# Topic Analysis of Superconductivity Literature by Semantic Non-negative Matrix Factorization

Valentin Stanev[1†,] Erik Skau[2], Ichiro Takeuchi[1], Boian S. Alexandrov[3]
[1]Department of Materials Science and Engineering, University of Maryland, College Park, MD, 20742, USA;
[2]Computer, Computational and Statistical Sciences Division, Los Alamos National Laboratory, Los Alamos, NM 87545, USA;
[3]Theoretical Division, Los Alamos National Laboratory, Los Alamos, NM 87545, USA;
[†] Corresponding author: vstanev@umd.edu

**Abstract.** We analyze a corpus consisting of more than 17,000 abstracts in the general field of superconductivity, extracted from the arXiv – an online repository of scientific articles. We utilize a recently developed topic modeling method called SeNMFk, extending the standard Non- negative Matrix Factorization (NMF) methods by incorporating the se- mantic structure of the text, and adding a robust system for determining the number of topics. With SeNMFk, we were able to extract coherent topics validated by human experts. From these topics, a few are relatively general and cover broad concepts, while the majority can be precisely mapped to specific scientific effects or measurement techniques. The topics also differ by ubiquity, with only three topics prevalent in almost 40% of the abstract, while each specific topic tends to dominate a small subset of the abstracts. These results demonstrate the ability of SeNMFk to produce a layered and nuanced analysis of large scientific corpora.

**Introduction**
Robust scientific activity is vital for economic and technological progress, and the ability to deal with various existing and emerging challenges. However, the current explosion of research work and publications is creating its own challenges. The total global research output already exceeds 2.6 million articles annually, and has grown at an average rate of 4% each year over the last decade[1]. This makes it impossible for individual scientists to keep up with important developments in their fields [12] and can overload entire journals[2]. Even the very recent research surge addressing the COVID-19 pandemic created a flood of papers, making it difficult to distinguish the important from trivial findings and thus impeding the efforts of scientists, health workers, and policymakers to quickly discover the most pertinent information[3]. This underscores the need for automated methods capable of organizing and analyzing the vast amounts of scientific publications appearing every day [1]. In the last two decades, Machine Learning (ML) methods for Natural Language Processing (NLP) have demonstrated a robust ability to model text data [22]. ML systems that can help to reduce the burden of analyzing existing literature will become an indispensable research

---

[1] https://ncses.nsf.gov/pubs/nsb20206/
[2] In 2018, the academic journal The Review of Higher Education had to suspend accepting submissions due to a two-year backlog.
[3] "Scientists are drowning in COVID-19 papers. Can new tools keep them afloat?", J. Brainard, Science (2020)

tool. However, many outstanding problems still have to be addressed before such systems become a reality. In every collection of documents, each document is a combination of some reoccurring themes or hidden (latent) concepts called topics. Extracting these topics and representing each document as their combination – known as topic modeling– is one of the most important and common tasks in the analysis of text corpora, and is a key in the efforts to reduce unorganized text corpora into actionable information. There are several classical topic modeling methods: Latent Semantic Analysis (LSA) [6], Probabilistic LSA (PLSA) [9], Latent Dirichlet Al- location (LDA) [4], Non-negative Matrix Factorization (NMF) [21]. These have been used extensively to analyze large text corpora in a variety of applications (see, for example, Refs. [10, 15, 7, 3, 20]).

Recently, we reported a new method, SeNMFk, which showed a superior performance in identifying the number of topics in several benchmark text corpora when compared to other state-of-the-art techniques [19]. SeNMFk extends the standard NMF methods in two key directions. Standard NMF models the topics by decomposing the term frequency-inverse document frequency (TF-IDF) representation of the corpus, a matrix X, into a product of two low-rank non- negative matrices: W, which represents the topics, and H, which represents the coordinates of each of the documents in the topic-space. SeNMFk utilizes a coupled minimization of the TF-IDF matrix, X, and the word-context matrix, M, to account for the semantic structure of the texts. Each cell of M represents the number of times two words (i, j) occur in a predetermined window (for example, +/- 5 words). The words in this window are the context words. Importantly, it has been demonstrated that word and context embeddings can be obtained by factorizing the normalized word-context matrix [13]. Second, SeNMFk includes a robust system for determining the latent dimension via random sampling and a subsequent custom clustering of the topic vectors. The custom clustering is used to determine the number of semantic-enhanced topics based on their stability.

Here, we extend the work of Ref. [19] by presenting a study of the performance of SeNMFk on a corpus of scientific publications: an important step in testing the ability of the model to extract valid coherent topics from a real-world data. We applied SeNMFk to a corpus of scientific abstracts from arXiv – an online repository of scientific articles[4]. We focused on abstracts from the field of superconductivity, which has been a subject of intense research ever since its discovery more than a century ago. This effort has led to many major discoveries and resulted in five Nobel prizes. Despite its long history, superconductivity remains a vibrant scientific field [8].

Via SeNMFk, we analyzed a corpus consisting of more than 17,000 ab-stracts. We were able to extract 29 stable topics which form 21 clusters. The topics were validated by human experts, and exhibit better coherence than pure NMF-extracted topics. The topics show a significant range in their specificity: a few relatively general topics cover broad concepts, while most of the topics can be precisely mapped to scientific effects, measurement techniques or applications. The topics also differ by ubiquity, with only three topics prevalent in almost 40% of the abstracts. Specific topics tend to dominate a small subset of the corpus – 1% - 4% of the abstracts. These results show the ability of SeNMFk to produce a nuanced analysis of large scientific corpora, and underscore the enormous potential of this method as a research tool.

**Methods: SeNMFk**

NMF is an unsupervised learning method that approximates a non-negative matrix, $X \in R_+^{F \times N}$, by a product of two factor matrices, $W \in R_+^{F \times k}$, and $H \in R_+^{k \times N}$, such that: $X_{ij} \approx \sum_{s=1}^{k} W_{is} H_{sj}$ [16], and W and H are both non-negative with one small dimension k. When NMF is applied to a text corpus for text mining, the basis patterns, represented by the columns of the matrix W, are the topics whose linear combinations span the entire corpus, and the total number of topics is equal to k. Each document is represented as a linear combination of the extracted topics with coefficients given by the columns of H.

---

[4] https://arxiv.org

There are a lot of notable topic modeling efforts within the NMF family of methods (see, for example, Refs. [5, 11]). However, a serious limitation of the classical NMF for topic modeling tasks is its inability to integrate the semantics of the words via, for example, word embeddings [14], successfully used in deep learning. Several recent works have presented different approaches for incorporating semantics information in NMF (see Refs. [2, 17, 18]) All these semantic-assisted NMF methods demonstrate an excellent coherence of the extracted topics and high-quality document clusters, but they require the number of latent topics, k, as a prior. In fact, accurately identifying the number of latent topics is a challenging task for all topic models. Yet, determining the correct latent dimension is imperative for a meaningful analysis of a corpus, as under- estimating the number of topics results in poor topic separation, under-fitting, while overestimation leads to noisy and superfluous topics, over-fitting.

To address this issue, recently we developed a robust topic modeling method called SeNMFk. SeNMFk is an NMF method that incorporates: (a) the semantic structure through term-context relations with a coupled minimization of the TF-IDF representation of the corpus, X, and Shifted Positive Point-wise Mutual Information (SPPMI) [13] matrix, M, and (b) determination of the latent dimension via random sampling of pairs of X and M matrices and a subsequent custom clustering of the topic vectors in W. This custom clustering allows us to find the latent dimension of the semantic-enhanced topics based on their stability. Here, we utilize SeNMFk, by solving the optimization problem, minimize

$$\underset{W\in\mathbb{R}_+^{F\times k}, H\in\mathbb{R}_+^{k\times N}, G\in\mathbb{R}_+^{k\times F}}{\text{minimize}} \frac{1}{2}\|X - WH\|^2 + \frac{\lambda}{2}\|M - WG\|^2,$$

which we solve efficiently by concatenating the TF-IDF matrix X with SPPMI matrix, M, and applying pyDNMFk[5] on the concatenation [X; M]. SeNMFk showed an excellent performance in identifying the number of topics in several benchmark text corpora, thus demonstrating its ability to avoid both under- and over-fitting [19].

**Data**

The corpus we analyzed consists of abstracts extracted from the arXiv – an open access repository of electronic preprints and postprints, currently containing more than 1.8 million scientific articles, predominantly from the fields of physics and other natural sciences. We collected the abstracts from the metadata repository for arXiv, made available through OAI-PMH (Open Archives Initiative Protocol for Metadata Harvesting). The abstracts were filtered by category and only the ones with "condmat.suprcon" ("Condensed Matter - Superconductivity") in their list of the categories were kept. There were more than 35,000 abstracts in this particular group. To ensure we are analyzing abstracts of a research screened by a robust peer-review process, we only considered abstracts which include DOI pointing to a published version of the article. To reduce the number of documents, we have focused on a span of twelve years — from 2007 to 2018. These years contain some significant developments in the field, including the discovery of an entire new high-temperature superconducting family and the emergence of topological superconductors as a major research topic. After applying these filters, there were 17,394 documents in the corpus, representing roughly half of all superconductivity abstracts in arXiv.

The raw abstract texts were preprocessed using standard NLP practices, including removal of the punctuation and standard English stop words, as well as the discovery of bigrams and trigrams. After the preprocessing, we converted the cleaned corpus to a TF-IDF representation. The vocabulary consists of 9433 tokens. It has to be noted that one significant factor complicating the analysis is the presence of chemical formulas. These obviously carry a lot of information but are not properly tokenized by standard NLP packages. Using specialized tools for chemical notation tokenization is an important future research direction.

---

[5]The software is publicly available at https://github.com/lanl/pyDNMFk

**Discussion**

SeNMFk extracted 29 topics; examining the top words in the topics, we can extract the themes they represent; the word cloud diagrams of a few selected topics are shown in Fig. 1[6]. The first important observation is that the topic specificity varies significantly, with some covering broad concepts, while others clearly being very thematically focused, reflecting, for example, a particular experimental technique or physical phenomenon. The second – connected – observation is that the relative importance of the topics also varies in a wide range: three of them are the leading topics in the respective 14%, 13%, and 12% of the abstracts (thus dominating a combined 39% of all documents). Another three topics form the next tier, leading in 8%, 8%, and 7%, respectively, of the abstracts (see Fig. 2a). None of the remaining 23 topics is leading in more than 5% of the abstracts, and eight topics are not leading in any of the abstracts. Not surprisingly, the three top topics are fairly broad.

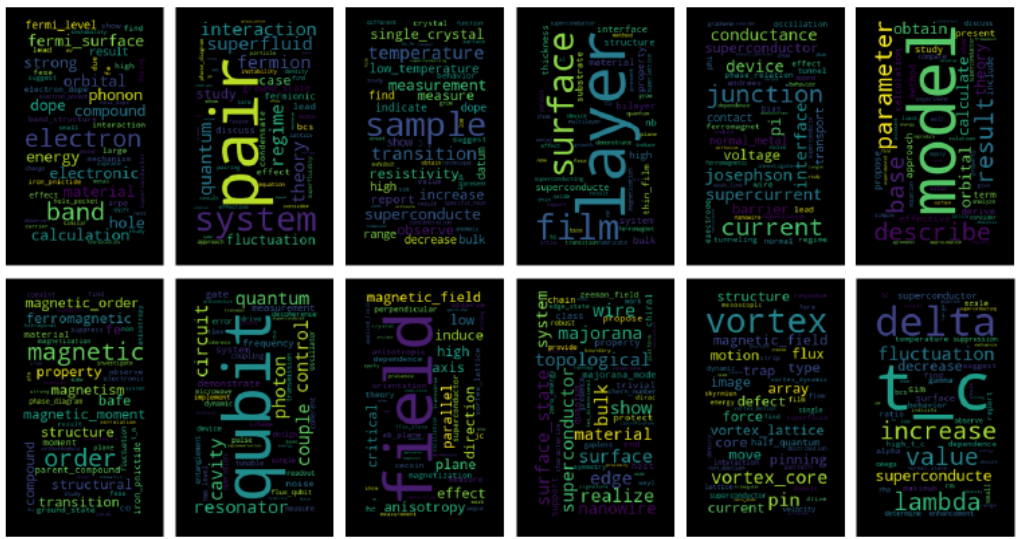

**Figure 1.** *Some of the topics discussed in the text, presented by their word clouds. The topics are ordered by their importance, from the most important at the top left to the least important at the bottom right: "Normal state", "Superconducting state", "Experiment – general", "Experiment – thin-film", "Experiment - Josephson junction", "Theory – general", "Magnetism", "Quantum computing", "Experiment - magnetic field", "Topological superconductivity", "Vortex", "Tc".*

One of them can be associated with a general de- scription of the properties and origin of the superconducting phase, with top words such as "superconducting pair", "superfluid", "phase diagram". Another one obviously stems from descriptions of the normal state electronic structure (band structure) underlying the superconducting phase, with words like "electron", "band", "fermi surface", "phonon" (electron-phonon interaction are the cause of superconductivity in many materials). The third of these broad topics is related to the experimental nature of many of the abstracts, with top words such as "sample", "measure", and "single crystal". The topics in the next tier are more specific. One contains words such as "layer", "surface", "film" and reflects the significant emphasis in the field on thin- film experiments. Another one clearly originates in work on Josephson junctions physics, with words like "junction", "josephson", "current", "conductance". The third topic in this tier is dominated by the words "result", "model", "calculate", "parameter", and "theory", and reflects the theoretical work in the field.

---

[6] All results are available at https://github.com/vstanev1/-NLP_arxiv_supercon

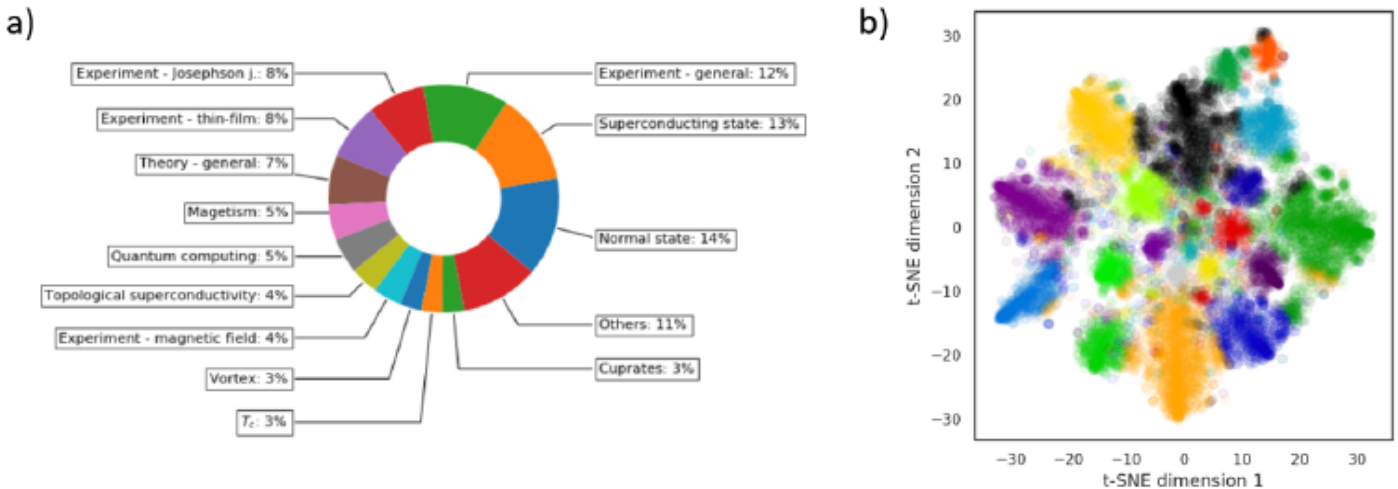

**Figure 2**. *a) The distribution of topics by the percentage of abstracts each topic dominates (note that category "Others" represents the combined contribution of the eight topics which dominate in less than 2% of the abstracts). b) The t-SNE projection of the abstracts using their coordinates in topic space, given by the columns of the H matrix.*

The remaining topics are typically very specific and rarely dominate in any given abstract. Two of the topics with appreciable weight (with top words "qubit" and "topological", and dominating 5% and 4% of the abstracts, respectively) reflect some of the most exciting recent developments in the field – the emergence of the idea of topologically nontrivial superconductors, and the development of superconducting devices for quantum computing. Other topics describe particular experimental techniques: one is associated with pressure studies (top words: "pressure", "gpa" = giga pascal) and another with applying external magnetic field (top words: "field", "magnetic field"). A few topics can be mapped to uses of superconductors for applications such as cables and digital circuits.

Interestingly, only two topics can be directly matched to a particular materials group. One topic (with top word "pseudogap") is based on studies of cuprates – the first high-temperature superconducting family. Another one is based on the second – iron-based – high-temperature superconducting family discovered in 2008 (one of the top words is "feas layer" – many of the materials in this group indeed contain FeAs layers). This surprising relative scarcity (all experimental work is done on concrete materials) is at least partially explained by the problems associated with tokenizing chemical names.

Extracting and analyzing the topics is only the first steps in the corpus analysis. To obtain a more nuanced general picture, we cluster the column-vectors of the H matrix obtained by the optimization procedure of SeNMFk (see Section 2).

This yields 21 well-separated clusters of abstracts formed around a single dominant topic, but containing appreciable contributions from other topics as well. For example, there is a cluster which is superposition of several theoretically- focused topics: the abstracts in this clusters with high probability present theoretical work. Another cluster is formed around the cuprate topic and contains almost equal components of the three most general topics (" superconducting state"," normal state"," experimental studies"), and noticeable contributions from few other specific topics – this reflects the diversity of approaches and techniques used in the field.

Using the results of SeNMFk, we can also visualize the entire corpus by projecting the coordinates of each abstract in topic-space using a dimensionality reduction technique. t-SNE was used in Fig. 2b, and the colors show the cluster index of the abstracts.

**Acknowledgments**
This research was funded by DOE National Nuclear Security Administration (NNSA) - Office of Defense Nuclear Nonproliferation R&D (NA-22), and supported by the LANL LDRD grant 20190020DR and DOE BES STTR DE- SC0021599.